\documentclass[10pt,a4paper,onecolumn]{article}
\usepackage{marginnote}
\usepackage{graphicx}
\usepackage{xcolor}
\usepackage{authblk,etoolbox}
\usepackage{titlesec}
\usepackage{calc}
\usepackage{tikz}
\usepackage{hyperref}
\hypersetup{colorlinks,breaklinks=true,
            urlcolor=[rgb]{0.0, 0.5, 1.0},
            linkcolor=[rgb]{0.0, 0.5, 1.0}}
\usepackage{caption}
\usepackage{tcolorbox}
\usepackage{amssymb,amsmath}
\usepackage{ifxetex,ifluatex}
\usepackage{seqsplit}
\usepackage{xstring}

\usepackage{float}
\let\origfigure\figure
\let\endorigfigure\endfigure
\renewenvironment{figure}[1][2] {
    \expandafter\origfigure\expandafter[H]
} {
    \endorigfigure
}

\usepackage{fixltx2e} 
\usepackage[
  backend=biber,
]{biblatex}
\bibliography{paper.bib}

\let\textttOrig=\texttt
\def\texttt#1{\expandafter\textttOrig{\seqsplit{#1}}}
\renewcommand{\seqinsert}{\ifmmode
  \allowbreak
  \else\penalty6000\hspace{0pt plus 0.02em}\fi}

\makeatletter
\let\href@Orig=\href
\def\href@Urllike#1#2{\href@Orig{#1}{\begingroup
    \def\Url@String{#2}\Url@FormatString
    \endgroup}}
\def\href@Notdoi#1#2{\def\tempa{#1}\def\tempb{#2}%
  \ifx\tempa\tempb\relax\href@Urllike{#1}{#2}\else
  \href@Orig{#1}{#2}\fi}
\def\href#1#2{%
  \IfBeginWith{#1}{https://doi.org}%
  {\href@Urllike{#1}{#2}}{\href@Notdoi{#1}{#2}}}
\makeatother

\newlength{\cslhangindent}
\newlength{\csllabelwidth}
\newenvironment{CSLReferences}[3] 
 {
  \setlength{\parindent}{0pt}
  \ifodd #1 \everypar{\setlength{\hangindent}{\cslhangindent}}\ignorespaces\fi
  \ifnum #2 > 0
  \setlength{\parskip}{#2\baselineskip}
  \fi
 }%
 {}
\usepackage{calc}

\usepackage[top=3.5cm, bottom=3cm, right=1.5cm, left=1.0cm,
            headheight=2.2cm, reversemp, includemp, marginparwidth=4.5cm]{geometry}

\titleformat{\section}
  {\normalfont\sffamily\Large\bfseries}
  {}{0pt}{}
\titleformat{\subsection}
  {\normalfont\sffamily\large\bfseries}
  {}{0pt}{}
\titleformat{\subsubsection}
  {\normalfont\sffamily\bfseries}
  {}{0pt}{}
\titleformat*{\paragraph}
  {\sffamily\normalsize}

\usepackage{fancyhdr}
\makeatletter
\let\ps@plain\ps@fancy
\fancyheadoffset[L]{4.5cm}
\fancyfootoffset[L]{4.5cm}

\definecolor{linky}{rgb}{0.0, 0.5, 1.0}

\newtcolorbox{repobox}
   {colback=red, colframe=red!75!black,
     boxrule=0.5pt, arc=2pt, left=6pt, right=6pt, top=3pt, bottom=3pt}

\newcommand{\ExternalLink}{%
   \tikz[x=1.2ex, y=1.2ex, baseline=-0.05ex]{%
       \begin{scope}[x=1ex, y=1ex]
           \clip (-0.1,-0.1)
               --++ (-0, 1.2)
               --++ (0.6, 0)
               --++ (0, -0.6)
               --++ (0.6, 0)
               --++ (0, -1);
           \path[draw,
               line width = 0.5,
               rounded corners=0.5]
               (0,0) rectangle (1,1);
       \end{scope}
       \path[draw, line width = 0.5] (0.5, 0.5)
           -- (1, 1);
       \path[draw, line width = 0.5] (0.6, 1)
           -- (1, 1) -- (1, 0.6);
       }
   }

\patchcmd{\@maketitle}{center}{flushleft}{}{}
\patchcmd{\@maketitle}{center}{flushleft}{}{}
\patchcmd{\@maketitle}{\LARGE}{\LARGE\sffamily}{}{}
\def\maketitle{{%
  
  \AB@maketitle}}
\makeatletter
\renewcommand\AB@affilsepx{ \protect\Affilfont}
\renewcommand\AB@affilnote[1]{{\bfseries #1}\hspace{3pt}}
\renewcommand{\affil}[2][]%
   {\newaffiltrue\let\AB@blk@and\AB@pand
      \if\relax#1\relax\def\AB@note{\AB@thenote}\else\def\AB@note{#1}%
        \setcounter{Maxaffil}{0}\fi
        \begingroup
        \let\href=\href@Orig
        \let\texttt=\textttOrig
        \let\protect\@unexpandable@protect
        \def\thanks{\protect\thanks}\def\footnote{\protect\footnote}%
        \@temptokena=\expandafter{\AB@authors}%
        {\def\\{\protect\\\protect\Affilfont}\xdef\AB@temp{#2}}%
         \xdef\AB@authors{\the\@temptokena\AB@las\AB@au@str
         \protect\\[\affilsep]\protect\Affilfont\AB@temp}%
         \gdef\AB@las{}\gdef\AB@au@str{}%
        {\def\\{, \ignorespaces}\xdef\AB@temp{#2}}%
        \@temptokena=\expandafter{\AB@affillist}%
        \xdef\AB@affillist{\the\@temptokena \AB@affilsep
          \AB@affilnote{\AB@note}\protect\Affilfont\AB@temp}%
      \endgroup
       \let\AB@affilsep\AB@affilsepx
}
\makeatother

\renewcommand\Affilfont{\sffamily\small\mdseries}
\ifnum 0\ifxetex 1\fi\ifluatex 1\fi=0 
  \usepackage[T1]{fontenc}
  \usepackage{lmodern}
  \usepackage[utf8]{inputenc}

\else 
  \ifxetex
    \usepackage{mathspec}
    \usepackage{fontspec}

  \else
    \usepackage{fontspec}
  \fi
  \defaultfontfeatures{Ligatures=TeX,Scale=MatchLowercase}

\fi
\IfFileExists{upquote.sty}{\usepackage{upquote}}{}
\IfFileExists{microtype.sty}{%
\usepackage{microtype}
\UseMicrotypeSet[protrusion]{basicmath} 
}{}

\usepackage{hyperref}
\hypersetup{unicode=true,
            pdftitle={A-SLOTH: Ancient Stars and Local Observables by Tracing Halos},
            pdfborder={0 0 0},
            breaklinks=true}
\let\addcontentslineOrig=\addcontentsline
\def\addcontentsline#1#2#3{\bgroup
  \let\texttt=\textttOrig\addcontentslineOrig{#1}{#2}{#3}\egroup}
\let\markbothOrig\markboth
\def\markboth#1#2{\bgroup
  \let\texttt=\textttOrig\markbothOrig{#1}{#2}\egroup}
\let\markrightOrig\markright
\def\markright#1{\bgroup
  \let\texttt=\textttOrig\markrightOrig{#1}\egroup}

\IfFileExists{parskip.sty}{%
\usepackage{parskip}
}{
\setlength{\parindent}{0pt}
\setlength{\parskip}{6pt plus 2pt minus 1pt}
}
\ifx\paragraph\undefined\else
\let\oldparagraph\paragraph
\renewcommand{\paragraph}[1]{\oldparagraph{#1}\mbox{}}
\fi
\ifx\subparagraph\undefined\else
\let\oldsubparagraph\subparagraph
\renewcommand{\subparagraph}[1]{\oldsubparagraph{#1}\mbox{}}
\fi

\title{A-SLOTH: Ancient Stars and Local Observables by Tracing Halos}

        \author[1, 2]{Mattis Magg}
          \author[3, 4, 5]{Tilman Hartwig}
          \author[1, 2]{Li-Hsin Chen}
          \author[4]{Yuta Tarumi}
    
      \affil[1]{Institut für Theoretische Astrophysik, Universität
Heidelberg, Germany}
      \affil[2]{International Max Planck Research School for Astronomy
and Cosmic Physics, University of Heidelberg (IMPRS-HD), Germany}
      \affil[3]{Institute for Physics of Intelligence, The University of
Tokyo, Japan}
      \affil[4]{Department of Physics, The University of Tokyo, Japan}
      \affil[5]{Kavli Institute for the Physics and Mathematics of the
Universe (WPI), The University of Tokyo, Japan}
  \date{\vspace{-7ex}}

\begin{document}
\maketitle

\marginpar{

  \begin{flushleft}
  \sffamily\small

  {\bfseries DOI:} \href{https://joss.theoj.org/papers/10.21105/joss.04417}{\color{linky}{10.21105/joss.04417}}

  \vspace{2mm}

  {\bfseries Software}
  \begin{itemize}
    \setlength\itemsep{0em}
    \item \href{https://github.com/openjournals/joss-reviews/issues/4417}{\color{linky}{Review}} \ExternalLink
    \item \href{https://gitlab.com/thartwig/asloth}{\color{linky}{Repository}} \ExternalLink
    \item \href{https://zenodo.org/record/6683682}{\color{linky}{Archive}} \ExternalLink
  \end{itemize}

  \vspace{2mm}

  \par\noindent\hrulefill\par

  \vspace{2mm}

  {\bfseries Editor:} \href{https://dfm.io/}{Dan Foreman-Mackey} \ExternalLink \\
  \vspace{1mm}
    {\bfseries Reviewers:}
  \begin{itemize}
  \setlength\itemsep{0em}
    \item \href{https://github.com/gregbryan}{@gregbryan}
    \item \href{https://github.com/kaleybrauer}{@kaleybrauer}
    \end{itemize}
    \vspace{2mm}

  {\bfseries Submitted:} 06 April 2022\\
  {\bfseries Published:} 26 June 2022

  \vspace{2mm}
  {\bfseries License}\\
  Authors of papers retain copyright and release the work under a Creative Commons Attribution 4.0 International License (\href{http://creativecommons.org/licenses/by/4.0/}{\color{linky}{CC BY 4.0}}).

    \vspace{4mm}
  {\bfseries In partnership with}\\
  \vspace{2mm}
  \includegraphics[width=4cm]{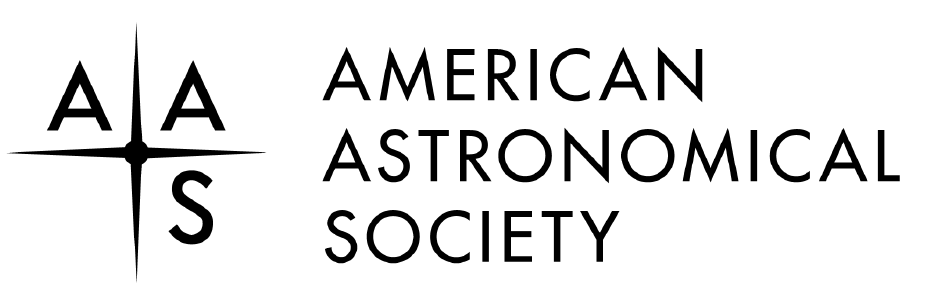}
  \vspace{2mm}
  \newline
  This article and software are linked with research article DOI \href{https://doi.org/10.3847/1538-4357/ac7150}{\color{linky}{10.3847/1538-4357/ac7150}}, published in the Astrophysical
Journal.
  
  \end{flushleft}
}

\hypertarget{summary}{%
\section{Summary}\label{summary}}

Galaxies are thought to reside inside of large gravitationally bound
structures of dark matter, so-called haloes. While the smallest of these
haloes host no or only a few stars, the biggest host entire clusters of
galaxies. Over cosmic history, haloes often collided and merged, forming
bigger and bigger structures. Merger trees, i.e., catalogues of haloes
evolving and connections between them as they grow and merge, have
become a vital tool in describing and understanding the history of
cosmological objects such as our Galaxy. Semi-analytical models, built
on top of such merger trees, are a common approach for theoretical
studies in cosmology. The semi-analytical nature of such models is
especially beneficial when the dynamic range in spatial and time scales
that need to be considered becomes too large for numerical simulations.

Ancient Stars and Local Observables by Tracing Halos (A-SLOTH) is such a
semi-analytical model and it is designed to simulate star formation in
the early Universe in a fast and accessible way. It uses merger trees,
either from numerical simulations or generated by statistical algorithms
to describe the history of galaxies. The processes of baryonic physics,
in particular gas cooling, star formation and stellar feedback are
described with approximations and statistical models. The range of
applications for this model is extensive and we, therefore, make it
available to the scientific community. We also provide
\href{https://a-sloth.readthedocs.io/en/latest/index.html}{full
documentation}.

\hypertarget{statement-of-need}{%
\section{Statement of need}\label{statement-of-need}}

Attempts to model the early Universe are notorious for the wide range in
spatial and time scales that they need to consider (Greif, 2015).
Therefore, a wide range of authors has resorted to developing
semi-analytical models for investigating the high-redshift Universe
(Brauer et al., 2019; de Bennassuti et al., 2017; Ishiyama et al., 2016;
Visbal et al., 2018). These models differ widely in their scope, and
often they are specifically geared towards addressing a specific issue
or question. With A-SLOTH, we offer a highly capable semi-analytical
model that can make predictions in numerous areas, ranging from 21-cm
cosmology to metal-poor stars in the Milky-Way (Chen et al., 2022;
Hartwig et al., 2015, 2018; Hartwig, Volonteri, et al., 2016; Hartwig,
Latif, et al., 2016; M. Magg et al., 2018, 2016; Mattis Magg et al.,
2022; Tarumi et al., 2020). This model was originally based on the
Extended-Press-Schechter algorithm and more specifically the GALFORM
code (Parkinson et al., 2008) but has since evolved to use merger trees
from numerical simulation. A-SLOTH has the primary purpose of quickly
simulating star formation in the early Universe and connecting it to
present-day and high-redshift observables, such as the metallicity
distribution function of the Milky Way or the ionization history of the
Universe. By modifying the parameters of the models governing the
high-redshift processes and repeated comparison of the results, a user
can investigate the relationship between assumptions about poorly
understood processes in the high-redshift Universe and astrophysical
observations. Because the model is highly optimized, A-SLOTH can build
up a Milky-Way-like galaxy star by star in only a few minutes. The code
is written in a modular way, such that users can add new physics and go
from an idea to a set of predictions in a short time and without
developing a completely new model of high-redshift star formation.

\hypertarget{acknowledgements}{%
\section{Acknowledgements}\label{acknowledgements}}

We appreciate the thorough testing and constructive feedback from Kaley
Brauer and Greg Bryan during code review. We thank Britton Smith for
very helpful advice with regards to the publication of open-source
software. We acknowledge funding from JSPS KAKENHI Grant Numbers
19K23437 and 20K14464, from Deutsche Forschungsgemeinschaft (DFG) via
the Collaborative Research Center (SFB 881, Project-ID 138713538) ``The
Milky Way System'\,' (sub-project A1), and from the
Max-Planck-Gesellschaft via the fellowship of the International Max
Planck Research School for Astronomy and Cosmic Physics at the
University of Heidelberg (IMPRS-HD).

\hypertarget{references}{%
\section*{References}\label{references}}
\addcontentsline{toc}{section}{References}

\hypertarget{refs}{}
\begin{CSLReferences}{1}{0}
\leavevmode\hypertarget{ref-Brauer2019}{}%
Brauer, K., Ji, A. P., Frebel, A., Dooley, G. A., Gómez, F. A., \&
O'Shea, B. W. (2019). {The Origin of r-process Enhanced Metal-poor Halo
Stars In Now-destroyed Ultra-faint Dwarf Galaxies}. \emph{Astrophysical
Journal}, \emph{871}(2), 247.
\url{https://doi.org/10.3847/1538-4357/aafafb}

\leavevmode\hypertarget{ref-chen22}{}%
Chen, L.-H., Magg, M., Hartwig, T., Glover, S. C. O., Ji, A. P., \&
Klessen, R. S. (2022). {Tracing stars in Milky Way satellites with
A-SLOTH}. \emph{arXiv e-Prints}, arXiv:2202.01220.
\url{http://arxiv.org/abs/2202.01220}

\leavevmode\hypertarget{ref-deBennassuti17}{}%
de Bennassuti, M., Salvadori, S., Schneider, R., Valiante, R., \&
Omukai, K. (2017). {Limits on Population III star formation with the
most iron-poor stars}. \emph{Monthly Notices of the Royal Astronomical
Society}, \emph{465}, 926--940.
\url{https://doi.org/10.1093/mnras/stw2687}

\leavevmode\hypertarget{ref-GreifReview}{}%
Greif, T. H. (2015). {The numerical frontier of the high-redshift
Universe}. \emph{Computational Astrophysics and Cosmology}, \emph{2}, 3.
\url{https://doi.org/10.1186/s40668-014-0006-2}

\leavevmode\hypertarget{ref-Hartwig15b}{}%
Hartwig, T., Bromm, V., Klessen, R. S., \& Glover, S. C. O. (2015).
{Constraining the primordial initial mass function with stellar
archaeology}. \emph{Monthly Notices of the Royal Astronomical Society},
\emph{447}, 3892--3908. \url{https://doi.org/10.1093/mnras/stu2740}

\leavevmode\hypertarget{ref-Hartwig16b}{}%
Hartwig, T., Latif, M. A., Magg, M., Bromm, V., Klessen, R. S., Glover,
S. C. O., Whalen, D. J., Pellegrini, E. W., \& Volonteri, M. (2016).
{Exploring the nature of the Lyman-{\(\alpha\)} emitter CR7}.
\emph{Monthly Notices of the Royal Astronomical Society}, \emph{462},
2184--2202. \url{https://doi.org/10.1093/mnras/stw1775}

\leavevmode\hypertarget{ref-Hartwig16a}{}%
Hartwig, T., Volonteri, M., Bromm, V., Klessen, R. S., Barausse, E.,
Magg, M., \& Stacy, A. (2016). {Gravitational waves from the remnants of
the first stars}. \emph{Monthly Notices of the Royal Astronomical
Society}, \emph{460}, L74--L78.
\url{https://doi.org/10.1093/mnrasl/slw074}

\leavevmode\hypertarget{ref-Hartwig18a}{}%
Hartwig, T., Yoshida, N., Magg, M., Frebel, A., Glover, S. C. O., Gómez,
F. A., Griffen, B., Ishigaki, M. N., Ji, A. P., Klessen, R. S., O'Shea,
B. W., \& Tominaga, N. (2018). {Descendants of the first stars: the
distinct chemical signature of second generation stars}. \emph{Monthly
Notices of the Royal Astronomical Society}.
\url{https://doi.org/10.1093/mnras/sty1176}

\leavevmode\hypertarget{ref-Ishiyama16}{}%
Ishiyama, T., Sudo, K., Yokoi, S., Hasegawa, K., Tominaga, N., \& Susa,
H. (2016). {Where are the Low-mass Population III Stars?}
\emph{Astrophysical Journal}, \emph{826}, 9.
\url{https://doi.org/10.3847/0004-637X/826/1/9}

\leavevmode\hypertarget{ref-Magg18}{}%
Magg, M., Hartwig, T., Agarwal, B., Frebel, A., Glover, S. C. O.,
Griffen, B. F., \& Klessen, R. S. (2018). {Predicting the locations of
possible long-lived low-mass first stars: importance of satellite dwarf
galaxies}. \emph{Monthly Notices of the Royal Astronomical Society},
\emph{473}, 5308--5323. \url{https://doi.org/10.1093/mnras/stx2729}

\leavevmode\hypertarget{ref-Magg16}{}%
Magg, M., Hartwig, T., Glover, S. C. O., Klessen, R. S., \& Whalen, D.
J. (2016). {A new statistical model for Population III supernova rates:
discriminating between {\(\Lambda\)}CDM and WDM cosmologies}.
\emph{Monthly Notices of the Royal Astronomical Society}, \emph{462},
3591--3601. \url{https://doi.org/10.1093/mnras/stw1882}

\leavevmode\hypertarget{ref-Magg21b}{}%
Magg, Mattis, Reis, I., Fialkov, A., Barkana, R., Klessen, R. S.,
Glover, S. C. O., Chen, L.-H., Hartwig, T., \& Schauer, A. T. P. (2022).
{Effect of the cosmological transition to metal-enriched star-formation
on the hydrogen 21-cm signal}. \emph{Monthly Notices of the Royal
Astronomical Society}. \url{https://doi.org/10.1093/mnras/stac1664}

\leavevmode\hypertarget{ref-Parkinson2008}{}%
Parkinson, H., Cole, S., \& Helly, J. (2008). {Generating dark matter
halo merger trees}. \emph{Monthly Notices of the Royal Astronomical
Society}, \emph{383}, 557--564.
\url{https://doi.org/10.1111/j.1365-2966.2007.12517.x}

\leavevmode\hypertarget{ref-Tarumi20}{}%
Tarumi, Y., Hartwig, T., \& Magg, M. (2020). {Implications of
Inhomogeneous Metal Mixing for Stellar Archaeology}. \emph{Astrophysical
Journal}, \emph{897}(1), 58.
\url{https://doi.org/10.3847/1538-4357/ab960d}

\leavevmode\hypertarget{ref-Visbal18}{}%
Visbal, E., Haiman, Z., \& Bryan, G. L. (2018). {Self-consistent
semi-analytic models of the first stars}. \emph{Monthly Notices of the
Royal Astronomical Society}, \emph{475}(4), 5246--5256.
\url{https://doi.org/10.1093/mnras/sty142}

\end{CSLReferences}

\end{document}